\documentclass[pra,twocolumn,showpacs,preprintnumbers,amsmath,amssymb,superscriptaddress]{revtex4-1}

\usepackage{graphicx}
\usepackage{dcolumn}
\usepackage{bm}
\usepackage{epsfig}
\usepackage{amsmath}
\usepackage{amssymb}
\usepackage{color}
\usepackage{bbm}
\usepackage{braket}
\usepackage[colorlinks=true,citecolor=blue,linkcolor=blue,urlcolor=blue]{hyperref}

\newcommand{\im}{\operatorname{i}\!}
\newcommand{\ex}{\operatorname{e}\!}
\newcommand{\id}{\operatorname{d}\!}

\begin{document}
 
%
\title{Ultrastrong coupling circuit QED in the radio-frequency regime}

\author{T. Jaako}
\email{tuomas.jaako@tuwien.ac.at}
\affiliation{Vienna Center for Quantum Science and Technology, Atominstitut, TU Wien, 1040 Vienna, Austria}

\author{J. J. Garc\' ia-Ripoll}
\affiliation{Instituto de F\' isica Fundamental, IFF-CSIC, Calle Serrano 113b, Madrid E-28006, Spain}

\author{P. Rabl}
\affiliation{Vienna Center for Quantum Science and Technology, Atominstitut, TU Wien, 1040 Vienna, Austria}

\date{\today}

\begin{abstract}
We study a circuit QED setup where multiple superconducting qubits are ultrastrongly coupled to a single radio-frequency resonator. In this extreme parameter regime of cavity QED the dynamics of the electromagnetic mode  is very slow compared to all other relevant timescales and can be described as an effective particle moving in an adiabatic energy landscape defined by the qubits. The focus of this work is placed on settings with two or multiple qubits, where different types of symmetry-breaking transitions in the ground- and excited-state potentials can occur. Specifically, we show how the change in the level structure and the wave packet dynamics associated with these transition points can be probed via conventional excitation spectra and Ramsey measurements performed at GHz frequencies. More generally, this analysis  demonstrates that state-of-the-art circuit QED systems can  be used to access a whole range of particle-like quantum mechanical phenomena beyond the usual paradigm of coupled qubits and oscillators. 
\end{abstract}

\maketitle

\section{Introduction}
Circuit quantum electrodynamics (QED) is a rapidly developing field where fundamental processes of quantum light-matter interactions are studied with  superconducting qubits (`artificial atoms') coupled to microwave resonators and transmission lines \cite{Blais2004,Wallraff2004,Gu2017}. Due to the extraordinary combination of strong coupling and very low losses, many quantum optical phenomena, such as vacuum Rabi splittings \cite{Wallraff2004}, photon blockade~\cite{Lang2011,Hoffman2011} or super- and subradiant decay \cite{mlynek2014}, have already been demonstrated in these systems with very high precision. Moreover, by using high-impedance resonators or by employing galvanic instead of electric coupling schemes, circuit QED systems can overcome fundamental bounds on the coupling strength in conventional cavity QED systems \cite{Devoret2007,debernardis2018}. It is then possible to access so-called ultrastrong (USC)  or deep-strong coupling regimes \cite{Ciuti2005,Casanova2010,forndiaz18,kockum19}, where the qubit-photon coupling is comparable to the photon energy and light-matter interactions become non-perturbative. These conditions  have recently been demonstrated  in experiments with superconducting qubits coupled to microwave resonators and transmission lines \cite{forndiaz10,baust16,forndiaz17,yoshihara17,Chen2017,Bosman2017}. When extended to multiple qubits, this regime could enable new applications such as protected quantum memories~\cite{nataff11}, ultrafast gate operations \cite{romero12} or entanglement harvesting schemes \cite{sabin2012,Armata2017}.

In regular cavity- and circuit QED systems with weak or moderate coupling, light-matter interactions are only effective close to resonance, where energy-conserving transitions between photons and atomic excitations can take place. This constraint does no longer apply in the USC regime, where photons and qubits with very dissimilar frequencies can still strongly influence each other. One specific limit of interest in this context is the low-mode-frequency or adiabatic regime \cite{Graham1984,Crisp1992,Liberti2006,liberti2006b,Ashhab2010}, where the bare oscillation frequency of the resonator mode, $\omega_r$, is much smaller than the qubit transition frequency, $\omega_q$. In this regime, the qubit state adjusts instantaneously to the slowly varying field amplitude and provides in turn an effective adiabatic potential for the photon mode. This separation of timescales is similar to the Born-Oppenheimer (BO) approximation in the description of nucleus-electron systems in molecular physics \cite{born1927}. In the weak-coupling limit, this physics finds applications, for example, for the readout of superconducting qubits or quantum dots through the off-resonant coupling to a low-frequency mode~\cite{johansson2006,nigg2009,lahay2009}. In the USC regime, this adiabatic picture is frequently employed in quantum optics and solid state physics to discuss symmetry-breaking effects in Rabi-, Dicke- and Jahn-Teller models, where the ground-state potential surface changes from a shape with a single minimum to a double-well or mexican-hat potential \cite{larson2017,bersuker2006}. In Refs.~\cite{Bakemeier2012,Ashhab2013,hwang2015,puebla2016,hwang2018} it has been shown in more detail that even for a single qubit this change in the adiabatic potential reproduces the properties of quantum-, excited-state- or dissipative phase transitions when $\omega_r\rightarrow 0$. However, reaching this regime with circuit QED systems requires electromagnetic modes in the radio-frequency regime where $\omega_r$ is only a few or a few tens of MHz. At these frequencies the mode is thermally occupied even at mK temperatures and electronic measurement techniques, which operate efficiently only above $\sim 1\,\mathrm{GHz}$, are not readily available. There is, nevertheless, experimental progress towards the quantum control of such modes. For example,  active ground state cooling and the readout of individual photon number states of a $ 170\,\mathrm{MHz} $ electromagnetic resonator has recently been demonstrated \cite{gely2019}.

In this paper we study the properties of circuit QED systems in the low-mode-frequency regime by considering a generic setup of two or multiple flux qubits coupled to a radio-frequency resonator. We derive an effective description of this system in terms of adiabatic BO potentials for the electromagnetic mode and discuss the characteristics of the resulting potential energy surfaces for different qubit numbers and coupling parameters.  Compared to related previous works~\cite{Graham1984,Crisp1992,Liberti2006,liberti2006b,Ashhab2010,larson2017,Bakemeier2012,Ashhab2013,hwang2015,puebla2016,hwang2018},  our main interest here is in the excited potential curves. These potentials can exhibit multiple first- and second-order symmetry-breaking transitions, even under conditions where the ground state potential still has a single minimum. This leads to a qualitatively new situation where properties of the symmetry-breaking transition  occurring at MHz frequencies can be probed via regular qubit-readout techniques operated in the GHz domain. As two specific examples, we describe how the change in the level structure and the dynamics of the wave packet splitting near the transition point can be detected via excitation spectra and Ramsey coherence measurements. This analysis demonstrates that quantum dynamics and phase-transition physics in the radio-frequency regime can be controlled and detected using state-of-the-art superconducting-circuit technology.

The remainder of the paper is structured as follows. After introducing in Sec.~\ref{sec:EDM} the basic model and its approximate treatment in the low-frequency limit, we provide in Sec.~\ref{sec:Potentials} a general overview of the characteristic features that can appear in the resulting adiabatic potential curves. In Sec.~\ref{sec:Spectrum} and Sec.~\ref{sec:Dynamics} we then discuss two schemes for detecting the symmetry-breaking transition in the excited potential. Finally, in Sec.~\ref{sec:Implementation} we perform a more rigorous justification for the single-mode approximation in a realistic setting and conclude our findings in Sec.~\ref{sec:Conclusions}.

\section{Circuit QED in the radio-frequency regime}\label{sec:EDM}

\begin{figure}[t]
    \begin{center}
        \includegraphics[width=\columnwidth]{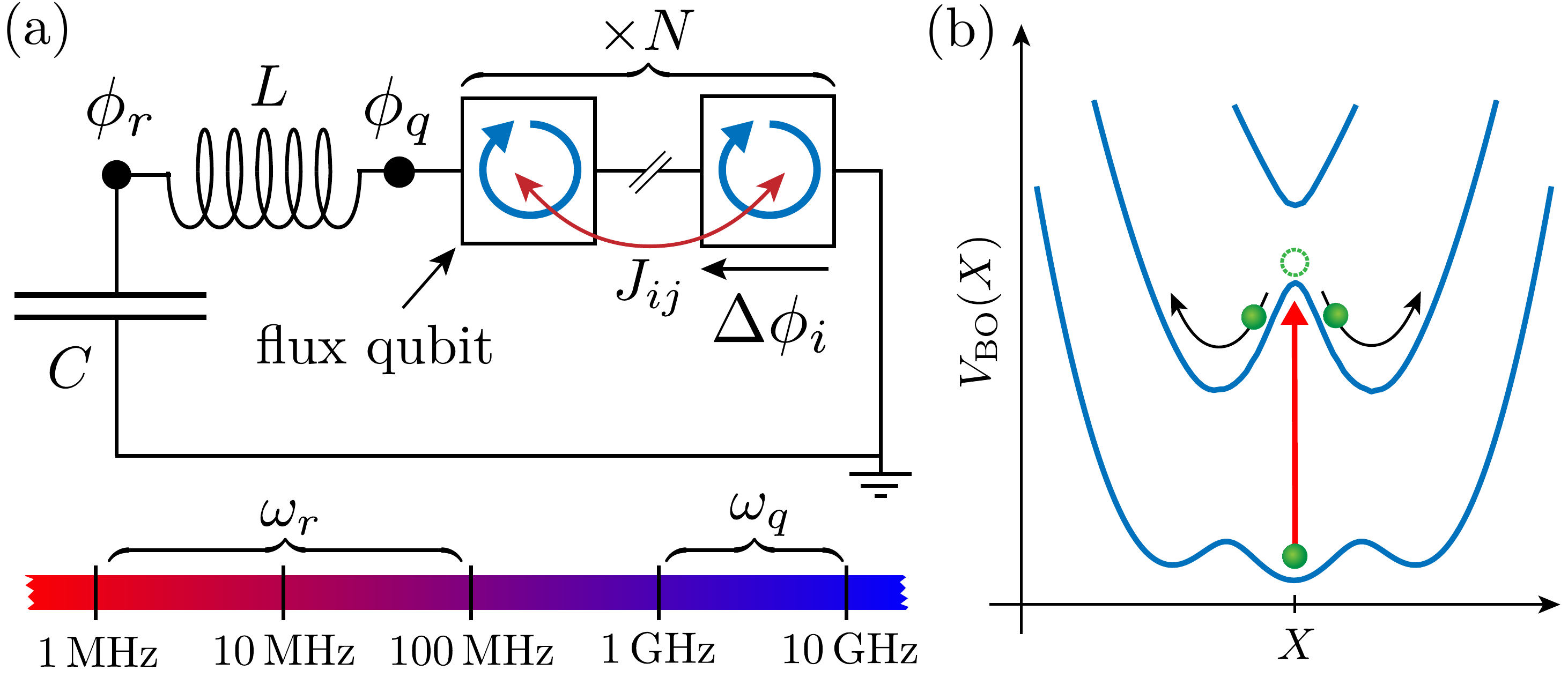}
        \caption{(a) Setup. Multiple superconducting flux qubits are coupled inductively to a single $LC $-resonator, giving rise to the collective qubit-photon coupling assumed in Eq.~\eqref{eq:h_cqed}. Additional direct qubit-qubit interactions $\sim J_{ij}$ can be engineered through the coupling to auxiliary SQUID loops, as described, for example, in Ref.~\cite{Plourde2004}. (b) In the radio-frequency regime, where the resonator frequency $\omega_r$ is much smaller than the qubit frequency, $\omega_q$, the dynamics of the electromagnetic mode can be modelled as an effective particle, which moves along the adiabatic BO potentials generated by the qubits. See text for more details.}
        \label{fig:1_setup}
    \end{center}
\end{figure}

We consider a circuit QED setup as shown in Fig.~\ref{fig:1_setup}, where $N$  flux qubits are coupled inductively to a lumped-element resonator with frequency $\omega_r$ and bosonic annihilation (creation) operator $a$ $(a^{\dagger})$. By taking only the two energetically lowest states $|g\rangle$ and $|e\rangle$ of each flux qubit into account, the quantized dynamics of this circuit is described by the general cavity QED Hamiltonian \cite{jaako2016,debernardis2018} ($ \hbar = 1 $)
\begin{equation}\label{eq:h_cqed}
\begin{split}
    H_{\rm cQED} = \omega_r a^{\dagger}a + &\sum_{i=1}^N \frac{\omega_q^i}{2} \sigma^i_z + \sum_{i=1}^N \frac{g_i}{2}(a^{\dagger} + a)\sigma^i_x\\
    +&  \sum_{i,j=1}^N \left( \frac{g_ig_j}{4 \omega_r} + J_{ij} \right)\sigma_x^i\sigma_x^j.
    \end{split}
\end{equation}
Here $\sigma_{x,y,z}$ are the usual Pauli operators and $\omega^i_{q}$ and $g_i$ denote the transition frequency and the coupling strength of each qubit. Apart from the collective qubit-photon coupling in the first line of Eq.~\eqref{eq:h_cqed}, this Hamiltonian also contains two direct qubit-qubit interaction terms. The so-called depolarization term $\sim g_ig_j$ represents the  gauge-dependent part of the qubit-photon interaction \cite{debernardis2018,debernardis2018b} and therefore should not be interpreted as a physical coupling between the qubits. In the current setup its origin can be understood as follows. When expressed in terms of the generalized flux variables indicated in Fig.~\ref{fig:1_setup}(a), the magnetic energy stored in the inductor reads
\begin{equation}\label{eq:Hmag}
H_{\rm mag} = \frac{(\phi_r - \phi_q)^2}{2L}.
\end{equation} 
Here $\phi_r\sim (a+a^\dag)$ is the oscillator flux variable  and $\phi_q=\sum_{i=1}^N \Delta \phi_i$, where the $\Delta \phi_i\sim \sigma_x^i$ represent the flux jumps across each qubit. Therefore, after expanding this energy in terms of the canonical variables, we obtain the photon-qubit interaction together with an apparent all-to-all coupling between the qubits~\cite{jaako2016,Armata2017}.
 
Equation~\eqref{eq:h_cqed} also includes additional direct qubit-qubit couplings with strengths $J_{ij}$.  Such couplings arise, for example, from the mutual inductance between nearby flux qubits, or, more generally, from a common coupling of the flux qubits to auxiliary superconducting quantum interference devices (SQUIDs), see, for example, Ref.~\cite{Plourde2004}.  In the latter case, the range, the sign and the strength of the elements $J_{ij}$ can be fully engineered and controlled by external bias currents. Although the presence of such qubit-qubit couplings is not essential for the main effects described in this work, they provide an additional tuning nob for the resulting potential surfaces discussed below.   

\subsection{Extended Dicke model}
For concreteness and to simplify the discussion, we will focus in the remainder of our analysis on the case of  identical qubits, $\omega_q^i=\omega_q$ and $g_i=g$, and all-to-all inter-qubit interactions, $J_{ij}= J$. In this case, Hamiltonian~\eqref{eq:h_cqed}  reduced to the extended Dicke model (EDM)~\cite{jaako2016,Todorov2014}
\begin{equation}\label{eq:h_edm}
\begin{split}
    H_{\rm EDM} =  \omega_r a^{\dagger}a + \omega_q S_z + g(a^{\dagger} + a)S_x + (1 + \varepsilon)\dfrac{g^2}{\omega_r}S_x^2,
    \end{split}
\end{equation}
where $S_{\alpha} = 1/2\sum_{i=1}^N \sigma_{\alpha}^{i}$ are collective spin operators. Here we have adopted the convention $J=\varepsilon g^2/(4\omega_r)$ \cite{debernardis2018}, such that the dimensionless parameter $\varepsilon$ characterizes the relative strength between qubit-qubit and qubit-photon interactions. We emphasize that none of the qualitative  findings of this work rely on the assumptions of identical qubits or purely collective couplings. We assume, however, that even in the presence of imperfections the model preserves its parity symmetry, i.e., it remains invariant under the transformation $\sigma_x^i\rightarrow - \sigma_x^i$ and $a\rightarrow - a$. For a more detailed derivation of the EDM for two basic circuit configurations, which correspond to $\varepsilon \geq 0 $ and $\varepsilon=-1$, the reader is referred to Refs.\ \cite{jaako2016,Armata2017} and \cite{Bamba2016}, respectively. Further details on the validity of the single-mode approximation assumed in our model are postponed to Sec.~\ref{sec:Implementation}.

\subsection{Born-Oppenheimer approximation}
In regular cavity and circuit QED setups the coupling $g$ between (artificial) atoms and photons is usually small and only resonant processes, where $\omega_r\approx \omega_q$, are relevant. In this regime, the physics of light-matter interactions is most conveniently described in terms of the bare photon number eigenstates, $|n\rangle$, where $a^\dag a|n\rangle=n|n\rangle$. Close to resonance, these photons can hybridize with matter excitations, which results, for example, in the appearance of a vacuum Rabi-splitting $\sim g\sqrt{N}$ in the excitation spectrum. In this work we are interested in a very different parameter regime, $\omega_r \ll \omega_q, g $ and $g^2/\omega_r\sim \omega_q$. Under such conditions the oscillation of the electromagnetic mode is slow compared to the qubit dynamics, while at the same time the coupling to the qubits has a substantial effect on the cavity and vice versa. Therefore, the representation of the electromagnetic field in terms of the bare photon number states is no longer useful in this regime. Instead, it is more convenient to describe the electromagnetic field as an effective massive particle moving in a set of adiabatic potential surfaces.

To model the static and dynamical properties of the circuit QED system in the limit $\omega_r\rightarrow 0$, we start by introducing the rescaled quadrature variables 
\begin{align}\label{eq:scaled_quadratures}
  X = \sqrt{\frac{\omega_r}{2\omega_q}}(a^{\dagger} + a),\qquad  P= \im\sqrt{\frac{\omega_q}{2\omega_r}}(a^{\dagger} - a),
\end{align}
which correspond to the usual position and momentum operators of a harmonic oscillator. After  normalizing all energies with respect to the qubit frequency, i.e., $\tilde H_{\rm EDM}=H_{\rm EDM}/\omega_q$, we obtain
\begin{align}\label{eq:h_P_Q}
   \tilde H_{\rm EDM}=  \frac{P^2}{2\mu } + \frac{X^2}{2} + \tilde H_{q}(X),
\end{align}
where $\mu=\omega^2_q/\omega^2_r
$ is the effective mass and 
\begin{align}\label{eq:h_adb}
   \tilde  H_q(X) = S_z + \sqrt{2}\lambda X S_x + (1 + \varepsilon)\lambda^2 S_x^2.
\end{align}
Here we have defined $ \lambda = \sqrt{g^2/(\omega_r\omega_q)} $ as the relevant dimensionless coupling parameter. 

The decomposition used in Eq.~\eqref{eq:h_P_Q} shows that for $\omega_r\ll \omega_q$ the `kinetic' energy term $\sim P^2$ is small compared to the potential and qubit energies. Thus, in direct analogy to the treatment of nucleus-electron systems in molecular physics, we can apply a BO approximation to separate the fast dynamics of the qubits and the slow motion of the resonator. Under this approximation the eigenstates of the combined  system are given by \cite{Liberti2006,liberti2006b}
\begin{align}\label{eq:adb_eig_state}
    \ket{\psi}_{n,k} = \int \id X \, \phi_{n,k}(X) \ket{X}\ket{\chi_n(X)},
\end{align}
where $ \ket{\chi_n(X)} $ is the adiabatic qubit eigenstate. It obeys the Schr\"odinger equation
\begin{align}\label{eq:adb_qubit_se}
   \tilde H_q(X)\ket{\chi_n(X)} = \tilde E_n(X)\ket{\chi_n(X)}
\end{align}
for a fixed value of $ X $. The dependence of $\tilde E_n(X)$ on the position coordinate provides an additional effective potential for the resonator wave function $\phi_{n,k}(X)$, which is a solution of the eigenvalue equation
\begin{align}\label{eq:adb_resonator_se}
    \left[ -\dfrac{1}{2\mu}\dfrac{\partial^2}{\partial X^2} + \tilde V_n(X) \right]\phi_{n,k}(X) = \tilde \epsilon_{n,k}\phi_{n,k}(X).
\end{align}
Here, $ \tilde V_n(X) = X^2/2 + \tilde E_n(X) $ is the total BO potential associated with the $n$-th qubit eigenstate. These adiabatic potentials are symmetric around $ X = 0 $ due to the invariance of Eq.~\eqref{eq:h_adb} under flipping the signs of $ X $ and $ S_x $ simultaneously.

The BO approximation neglects couplings between the adiabatic energy eigenstates, which are induced by the momentum operator. The main off-diagonal correction terms, which couple Eq.\ \eqref{eq:adb_resonator_se} for different $ n $, are of the form
\begin{equation}
  C_{n,m}=\frac{1}{\mu} \frac{\partial \phi_{n,k}}{\partial X} \frac{\langle \chi_m(X)| \sqrt{2}\lambda S_x | \chi_n(X)\rangle}{ \tilde{E}_{n}(X) - \tilde{E}_m(X)},
\end{equation}
where we used the relation $\langle \chi_m(X)| \frac{\partial}{\partial_X} |\chi_n(X)\rangle= \langle \chi_m(X)| \frac{\partial \tilde H_q(X)}{\partial X}|\chi_n(X)\rangle/[\tilde{E}_{n}(X) - \tilde{E}_m(X)]$. For moderate couplings and electromagnetic eigenstates near the potential minimum, where $\frac{\partial \phi_{n,k}}{\partial X}\sim \sqrt{\mu}$, we obtain a scaling $C_{n,m}\sim \lambda/\sqrt{\mu}$. In all the numerical examples discussed in this work we consider the parameter regime $\mu=10^4-10^6$, where the adiabatic  condition  $|C_{n,m}|\ll 1$  is well-satisfied for the relevant potential curves. However, as discussed in Sec.~\ref{subsec:General} below, for large couplings, $\lambda>1$, some of the excited potential surfaces are only separated by higher-order avoided crossings. In this case, $|\tilde{E}_{n}(X) - \tilde{E}_m(X)|\ll 1$ and corrections beyond the BO approximation may become relevant.

\section{Adiabatic potential surfaces}\label{sec:Potentials}
The energy landscape formed by all the $V_n$ is fully determined by the qubit Hamiltonian $H_q(X)$ and depends on the  interaction parameters $\lambda$ and $\varepsilon$ as well as the number of qubits, $N$. In this section we summarize  the characteristic features of these potentials in different parameter regimes.

\begin{figure}[t]
    \begin{center}
        \includegraphics[width=\columnwidth]{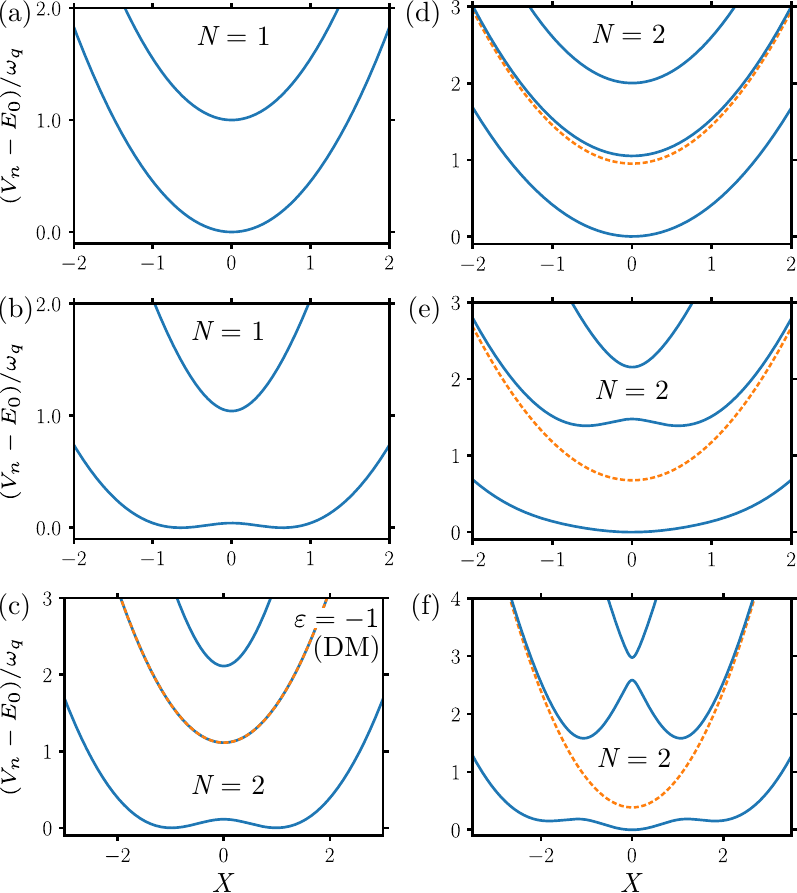}
        \caption{(a-b) Plot of the BO potentials for the Rabi model ($N=1$) for a coupling parameter (a) below, $ \lambda^2 = 0.1 $, and (b) above, $ \lambda^2 = 1.5 $, the ground-state instability. In (c) the corresponding  potentials are shown for the DM [$ \varepsilon = -1 $ in Eq.~\eqref{eq:h_edm}] for $ N = 2 $ qubits and for $ \lambda^2 = 0.8 $, and in (d-f) for two non-interacting qubits ($\varepsilon=0$) and (d) $ \lambda^2 = 0.1 $, (e) $ \lambda^2 = 0.8 $, and (f) $ \lambda^2 = 2.1 $. For the cases with $N=2$ qubits, the solid blue lines represent the triplet states and the dashed orange line the singlet potential. All energies are plotted with respect to the minimum of the lowest potential curve, $E_0={\rm min}\{V_{n=0}(X)|X\}$.}
        \label{Fig:2_BOPotentials}
    \end{center}
\end{figure}

\subsection{Ground-state symmetry breaking}

In Fig.~\ref{Fig:2_BOPotentials}(a) and (b) we first plot  $V_{n=0,1}(X)$ for the simplest case of a single qubit, where $H_{\rm EDM}$ reduces to the quantum Rabi model. For $\lambda\ll1$ the original harmonic potentials  $\tilde V_{0,1}(X)\simeq X^2/2 \pm 1/2$ are only weakly perturbed and have a single minimum at $X_{\rm min}=0$. For increasing coupling strength the ground state potential becomes shallower until it transitions  into a double-well potential with two degenerate minima at $X_{\rm min} \approx \pm \sqrt{\lambda^4 - \lambda_c^4}/(\sqrt{2}\lambda\lambda_c^2)$ above the critical coupling $\lambda_c=1$. As indicated in Fig.~\ref{Fig:2_BOPotentials}(c) for $N=2$, the same qualitative behavior is also found for the $N$-qubit Dicke model (DM), which corresponds to the case $\varepsilon=-1$ in the current notation. In this specific situation the potentials have the simple analytic form \cite{Liberti2006,larson2017}
\begin{equation}\label{eq:adb_pot_dm}
    \tilde V_{s,m_z}^{\rm (DM)}(X) = \dfrac{X^2}{2} + m_z\sqrt{1 + 2\lambda^2 X^2},
\end{equation}
and can be labelled by the total spin quantum number $s$ and the spin projection $m_z = -s, s + 1, \dots, s - 1, s$ associated with the qubit states $|\chi_{s,m_z}(X=0)\rangle$. This result shows that also for larger $N$ the transition occurs first in the ground-state potential, $m_z=-N/2$, but at a reduced coupling parameter $\lambda_{c}=1/\sqrt{N}$.

This change from a single-well to a double-well structure of the adiabatic potential is familiar from studies of Rabi-, Dicke- and Jahn-Teller-type models, where even at larger $\omega_r$ this picture explains the observed symmetry breaking in the ground state. Here, symmetry breaking means that for $\lambda>\lambda_c$ the tunnel-splitting between the two lowest resonator states, $\Delta \epsilon$, is exponentially suppressed such that in a realistic setting any weak perturbation will randomly localize the system in one of the degenerate minima. Specifically, by approximating the ground-state potential $V_{N/2,-N/2}^{({\rm DM})}(X)$ by an equivalent quartic double-well potential with the same minima and the same barrier height, we obtain (see e.g. Ref.~\cite{Garg2000}) 
\begin{equation}
    \Delta \tilde \epsilon |_{\lambda>\lambda_c} \approx \dfrac{8}{\sqrt{\pi}}A \ex^{-S_0},
\end{equation}
with an exponent $S_0 = \frac{2}{3}\sqrt{\mu (\lambda^2 - \lambda_c^2)^3(\lambda^2+\lambda_c^2)}/(\lambda^2\lambda_c^4)$ and prefactor $ A = \sqrt[4]{(\lambda^2 - \lambda_c^2)^5/(\mu(\lambda^2 + \lambda_c^2))}/(\lambda\lambda_c^2) $. This means that for $\lambda\gg \lambda_c$ the tunneling is suppressed by $\Delta \tilde{\epsilon}\sim \ex^{-2\sqrt{\mu}\lambda^2N^2/3}$. When passing from the single-well to the double-well configuration the potential becomes purely quartic at $\lambda=\lambda_c$, in which case the minimal energy splitting is  \cite{vranicar2000}
\begin{equation}
    \Delta \tilde \epsilon |_{\lambda=\lambda_c}=  c \sqrt[3]{\dfrac{\lambda_c^2}{\mu^2}},
\end{equation}
with a numerical prefactor of $c\approx 1.1$. This means that the density of states at the transition point, $\tilde \nu_{\lambda=\lambda_c}= 1/\Delta \tilde \epsilon$,  scales as $\sim \sqrt[3]{N\mu^2}$ and diverges in the classical limit $\mu\rightarrow \infty$. Note, however, that when compared to the density of states of the unperturbed harmonic potential, $\tilde \nu_{\lambda=0}=\sqrt{\mu}$, the peak at the transition point is not very pronounced. This illustrates the necessity to use either many qubits or very large ratios $\mu=\omega_q^2/\omega_r^2$ to detect sharp experimental signatures associated with this quasi-divergence. It has been estimated that the regime $\mu>10^4$ can be reached using effective implementations of the Rabi model in trapped ion systems~\cite{Puebla2017}. In circuit QED, reaching this parameter regime requires resonator frequencies below $100$ MHz.

\subsection{Non-interacting qubits}
While the DM has been the subject of many theoretical studies, it only represents a very special class of cavity and circuit QED setups with strong ferroelectric interactions~\cite{debernardis2018}.  In Fig.~\ref{Fig:2_BOPotentials}(d-f) we now show the potential curves for another relevant example of two non-interacting (or only weakly interacting) qubits, where  $\varepsilon\simeq 0$. In this case we see that  as the coupling parameter $\lambda$ increases, the absolute minimum of the ground state potential remains at $X_{\rm min}=0$, while for $\lambda \gtrsim 1$ two additional local minima appear around $X\approx \pm 2$. More important for the current work, already at an intermediate coupling strength of $\lambda=1/\sqrt{2}$ a symmetry-breaking transition occurs in the first excited potential, see App.\ \ref{app:per_thy}. As explained in more detail in Secs.~\ref{sec:Spectrum} and \ref{sec:Dynamics} below, this has important practical implications. The symmetry-breaking effect now occurs at an absolute frequency scale set by the qubit frequency $\sim \omega_q$, at which efficient electronic readout techniques are readily available.

Compared to the DM,  another  qualitative difference is that for $\varepsilon=0$ (as well as for any $\varepsilon \neq -1$) the degeneracy between potential curves of different spin quantum numbers is lifted by the $S_x^2$ term.  For example, for $N=2$ we obtain a splitting of (see App.~\ref{app:per_thy}),
\begin{equation}\label{eq:DeltaE}
\Delta \tilde E(X=0) = (1+\varepsilon)\lambda^2
\end{equation}
between the triplet and singlet potentials at $X=0$. Therefore, even though the oscillator coordinate is zero, there is still an energy penalty for the triplet state compared to the singlet. This somewhat counterintuitive result can be understood by looking at the original  magnetic interaction Hamiltonian 
given in Eq.~\eqref{eq:Hmag}. At $ X = 0 $, i.e. $ \phi_r = 0 $, there will be an energy cost for states for which $ \phi_q \neq 0 $, such as the triplet state. For the singlet this contribution vanishes.

\subsection{General structure}\label{subsec:General}

\begin{figure}[t]
    \begin{center}
        \includegraphics[width=\columnwidth]{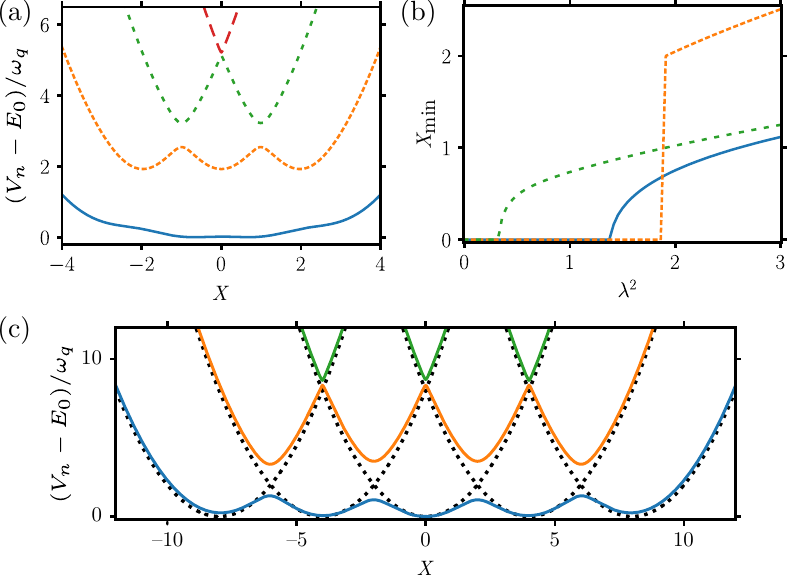}
        \caption{(a) BO potentials for a system of three qubits and $\varepsilon=0.02$. At the chosen coupling strength of $ \lambda^2 \approx 1.86 $, this system exhibits a first-order transition with three degenerate potential wells. (b) Location of the minima of the BO potentials shown in (a) as a function of $ \lambda $. In this plot we have omitted the highest excited state since its minimum is always located at zero due to symmetry. There is always another degenerate minima located at $ -X_{\rm min} $. (c) Plot of the three lowest BO potentials for a four-qubit system for $ \lambda \gg 1 $ and $\varepsilon=0$. The black dotted lines indicate the zeroth-order potentials given in Eq.~\eqref{eq:ZerothOrderPot}. See also Ref.~\cite{jaako2016}. Note that for better visibility, all the plots only show the potential curves for qubit states with maximal spin $s=N/2$.}
        \label{fig:3_bo_pots_multiqubit}
    \end{center}
\end{figure}

For a larger number of qubits the energy potential landscapes become considerably more involved and depending on $\lambda$ and $\varepsilon$ the individual potentials curves can exhibit various local and global minima. This is illustrated in Fig.~\ref{fig:3_bo_pots_multiqubit}(a) and (b) for the example of $N=3$ and $\varepsilon= 0.02$.  Apart from the formation a double-well structure in the ground- and second excited potential curve, in this case  we also obtain  a triple-well potential  with three degenerate minima at a value of $\lambda^2 \simeq 1.86$. This situation corresponds to a first-order phase transition point, where the potential at $X=0$ remains stable, but the two outer wells become lower in energy after the transition point [cf. Fig.~\ref{fig:3_bo_pots_multiqubit}(b)]. 

For a fully symmetric system the total number of distinct potential curves scales as $(N/2 + 1)^2$ (for an even number of qubits) and as $2^N$, if different parameters for each qubit are taken into account. To obtain a better intuition about the basic potential landscapes that can arise, it is instructive to consider the limit of  very large coupling, $\lambda \gg 1$, following the analysis presented in Ref.~\cite{jaako2016}. In this limit, the terms $\sim S_x$ and $\sim S_x^2$  dominate over the bare qubit splitting $\sim S_z$. By neglecting the contribution of the $S_z$ term completely, the eigenstates $|\chi^0_{s,m_x}(X)\rangle$ are also eigenstates of $ S_x $ and can be labelled by the total spin $s$ and its projection along the $x$-axis, $m_x$.  The corresponding zeroth-order potential curves are
\begin{align}\label{eq:ZerothOrderPot}
    \tilde  V_{s,m_x}^{(0)}(X) = \dfrac{1}{2} \left(X + \sqrt{2}\lambda m_x\right)^2   + \varepsilon \lambda^2 m_x^2,
\end{align}
which are just a set of displaced parabolas indicated by the dashed lines in Fig.~\ref{fig:3_bo_pots_multiqubit}(c). As detailed in Ref.~\cite{jaako2016} and App.~\ref{app:per_thy}, the presence of the $S_z$ term in the full model leads to a coupling between the states with $m_x$ and $m_x\pm 1$. Focusing on the limit of very small values for $\varepsilon$, this coupling lowers, first of all, each well by  $\Delta \tilde V_{s,m_x}= (m_x^2 - s(s + 1))/(2\lambda^2)$, due to second-order couplings to energetically higher potentials. Further, at the crossing points $ X \simeq -\lambda(2m_x + 1)/\sqrt{2} $, the $S_z$ terms lifts the degeneracy and leads to an avoided crossing with a splitting of $\tilde \Delta_{m_x,m_x + 1} = \sqrt{s(s+1)-m_x(m_x+1)}$. As indicated in Fig.~\ref{fig:3_bo_pots_multiqubit}(c), the actual adiabatic potentials  $V_n(X)$ are  the resulting connected potential curves. Note that in this picture, which holds for  $\lambda\gg1$, the excited  potentials are only separated through higher-order avoided crossings. For general $\lambda$, the potential energy surface can then be understood as a smooth interpolation between the stack of parabolas $\tilde V_{s,m_z}(X)= m_z+X^2/2$ for the uncoupled system into the degenerate-well structure depicted in Fig.~\ref{fig:3_bo_pots_multiqubit}(c).

\subsection{Weak-coupling limit}
Finally, let us briefly remark on the limit $\lambda\ll 1$, where the system is far away from the instability points. In this case we can expand the adiabatic potential to second order in $\lambda$ to obtain a state-dependent shift of the oscillation frequency, $ \omega_{r}^{m_z} = \omega_r + m_z \delta \omega_r$  with  $ \delta \omega_r =  g^2/(2\omega_q) $. As long as $\delta \omega_r<\omega_r$, we can make a rotating wave approximation and describe this shift in terms of an effective Hamiltonian 
\begin{align}\label{eq:HStark}
    H_{\rm eff} &\simeq \omega_r a^\dag a +  \omega_q S_z+\dfrac{g^2}{4\omega_q}( 2a^\dag a +1)S_z \\
    &\quad+ \left[ (1 + \varepsilon)\dfrac{g^2}{2\omega_r} - \dfrac{g^2}{4\omega_q} \right] (\vec S^2 - S_z^2),\nonumber
\end{align}
which corresponds to the usual Stark-shift Hamiltonian for a far detuned cavity QED system. The third term in the first line of Eq.~\eqref{eq:HStark} represents a shift of the qubit frequency proportional to the cavity photon number. This type of interaction is frequently encountered in measurement schemes for quantum dots or superconducting qubits~\cite{johansson2006,nigg2009,lahay2009,Gu2017} and has been exploited in Ref.~\cite{gely2019} to resolve individual photon number states of a $ 170\,\mathrm{MHz} $ resonator~\cite{footnote}. At slightly larger couplings or even lower resonator frequencies, where $\delta \omega_r\gtrsim \omega_r$, the photon number is no longer a conserved quantity and this picture breaks down.

\section{Excitation spectrum}\label{sec:Spectrum}
In  regular circuit QED systems both $\omega_r$ and $\omega_q$ are in the GHz regime and both the qubit and the resonator mode can be probed using efficient electronic readout techniques.  In the RF domain this is no longer possible using conventional methods and therefore all the properties of the electromagnetic mode must be inferred through measurements performed on the qubits. As a first example, we consider in this section the single-qubit excitation spectrum $S(\omega)$, which is proportional to the excited state population of the first qubit when being driven by a weak external field of frequency $\omega\sim \omega_q$. The excitation spectrum is then given by
\begin{equation}\label{eq:Spectrum}
    S(\omega) = \dfrac{\Gamma}{2}{\rm Re} \int_0^\infty \id\tau\, \langle \sigma_x^1(\tau)\sigma_x^1(0)\rangle \ex^{\im\omega \tau},
\end{equation}
where the expectation value is taken with respect to the equilibrium state of the systems in the absence of the driving field. In Eq.~\eqref{eq:Spectrum} we have  introduced a characteristic decay rate $ \Gamma$, with which the correlation function decays. Since in the parameter regime of interest the qubits are not strongly perturbed by the coupling to the resonator mode, $\Gamma$ can simply be taken as the bare decay rate of the excited qubit state. In the numerical examples discussed in this section we assume moderate values of $\Gamma\sim\omega_r$, which is still large compared to decoherence rates achieved with state-of-the-art flux qubits~\cite{Yan2016}.

\subsection{Mode splitting in the low-frequency regime} 
In a first step we assume that the combined system is prepared close to the absolute ground state $\ket{\rm GS} $ (for example by actively cooling the resonator mode \cite{gely2019}). In this case the excitation spectrum is given by
\begin{align}\label{eq:lin_resp}
    S(\omega) = \sum_f \dfrac{1}{4}\dfrac{\Gamma^2|\langle f| \sigma_x^1 \ket{\rm GS}|^2}{(\omega - \omega_{f})^2 + \Gamma^2/4},
\end{align}
where  the sum runs over all final states $ \bra{f} $, which are separated by a frequency $\omega_f$ from the ground state.

\begin{figure}[t]
    \begin{center}
        \includegraphics[width=\columnwidth]{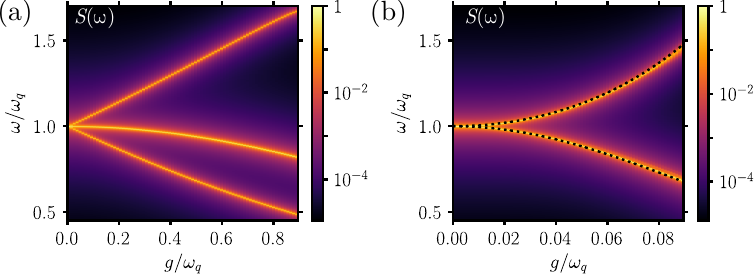}
        \caption{Spectrum $S(\omega)$ of the EDM for $N=2$ qubits and (a) on resonance, $\omega_r=\omega_q$, and (b) for $\omega_r/ \omega_q=0.01$. The black dotted lines in (b) indicate the energy of the triplet and singlet potential, relative to the ground state and at $ X = 0 $. Note that for the parameters assumed in (b) the symmetry-breaking transition occurs at a value of $g/\omega_q \simeq 0.07$.  For both plots we have used a fixed decay rate $ \Gamma/\omega_q = 0.005$ for all excited qubit states.}
        \label{fig:4_spectrum_disp_res}
    \end{center}
\end{figure}

In Fig.~\ref{fig:4_spectrum_disp_res} (a) and (b) we first compare the excitation spectrum of a conventional, i.e., resonant cavity QED system with the case of a circuit QED system in the low-frequency regime. For both plots $N=2$ and we use the more common convention to plot the spectrum as a function of the coupling strength $g$ instead of $\lambda$. In this case the resonant system exhibits the familiar Rabi splitting $\sim \sqrt{2} g$, which arises from the hybridization of the photon with the triplet state $|T\rangle$. This splitting is visible for $\sqrt{2}g>\Gamma$, which defines the strong coupling regime for a resonant cavity QED system. The additional resonance in the middle is the singlet state $|S\rangle$. The singlet state is decoupled from the cavity mode, but its energy relative to the ground state still decreases for larger coupling strengths.

In the radio-frequency regime, where $\omega_r/\omega_q=0.01$, the mode frequency is no longer resolved on the scale of this plot. However, the cavity mode has still a drastic influence on the spectrum through the induced splitting between the triplet and the singlet state [see Eq.~\eqref{eq:DeltaE}]. To be observable, this splitting must exceed the qubit decay rate $\Gamma$ and therefore we identify 
\begin{equation}
    g > \sqrt{\Gamma \omega_r}
\end{equation} 
as the minimal strong-coupling condition for the low-frequency regime.

\subsection{Probing the symmetry-breaking transition}
The dashed lines in Fig.~\ref{fig:4_spectrum_disp_res}(b) indicate the excitation frequencies derived from the energies $ V_n(X=0)$. The perfect agreement demonstrates that the dominant peaks of $S(\omega)$ at low temperatures probe the adiabatic potentials curves at $X=0$. This can be understood in terms of the BO wave function in Eq.~\eqref{eq:adb_eig_state} and the fact that for small and moderate couplings the qubit states $|\chi_n(X)\rangle$ are only weakly dependent on $X$. The excitation spectrum can then be approximated as 
\begin{align}\label{eq:SpectrumApp}
    S(\omega) \simeq C \sum_k\dfrac{\Gamma^2|\langle \phi_{1,k} | \phi_{0,0} \rangle|^2}{(\omega - \omega_{k0})^2 + \Gamma^2/4},
\end{align}
with a constant prefactor $C=|\bra{\chi_{1}(0)} \sigma^1_x \ket{\chi_0(0)}|^2/4$ and $\omega_{k0} = \epsilon_{1,k} -  \epsilon_{0,0}$. We see that the spectrum is mainly determined by excited states with a big overlap with the ground-state wave function $\phi_{0,0}(X)$, which only extends over a scale $\sim 1/\sqrt{\mu}$ around $X=0$. As a result, the spectral lines shown in Fig.~\ref{fig:4_spectrum_disp_res}(b) exhibit no particular feature at the transition point $\lambda^2=0.5$, which corresponds to the value of $g/\omega_q \simeq 0.07$ in this plot. 

To identify spectral signatures of the symmetry-breaking transition in the excited state, Figs.~\ref{fig:5_spectrum_thermal}(a) and (b) now show a zoom of the triplet line for  $\lambda \geq \lambda_c$ together with the lineshapes of $S(\omega)$ for different values of $\lambda$ below and above the transition point. For $\lambda<\lambda_c$ we observe a small decrease of the height of the resonance peak, which can be mainly attributed to the decrease in $C$, i.e., the change in the qubit transition matrix element. However, at $\lambda=\lambda_c$ there is an additional sudden drop in the height of the resonance, which for $\lambda >\lambda_c$ also gets substantially broader. This suppression can again be understood from the overlap between $\phi_{0,0}(X)$ and the few lowest wave functions in the excited state potential, which become considerably broader at the transition point. For sufficiently small $\Gamma$, the appearance of additional sidebands below the main spectral line indicate the excitation of motional states located in one of the two displaced potential wells.

\subsection{Spectrum at finite temperature}\label{subsec:spectrum}
\begin{figure}[t]
    \begin{center}
        \includegraphics[width=\columnwidth]{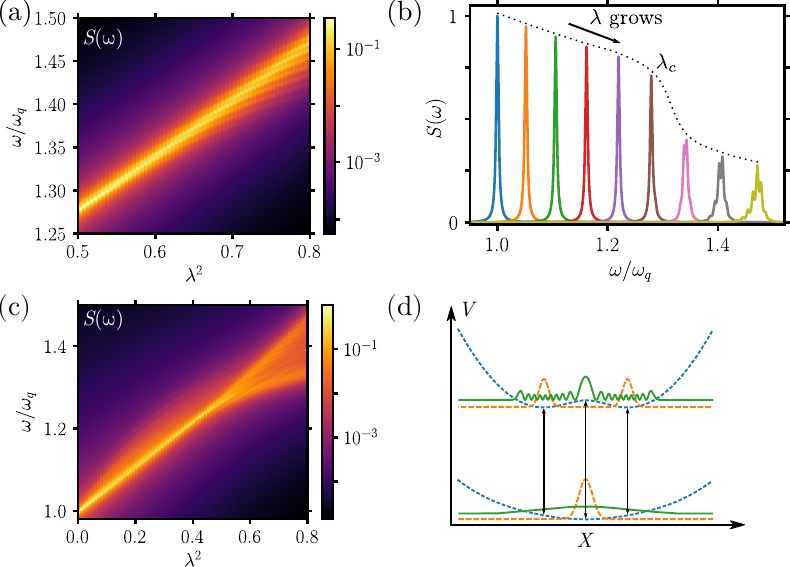}
        \caption{(a) Zoom of the triplet line of the spectrum $S(\omega) $ shown in Fig.~\ref{fig:4_spectrum_disp_res} (b) for coupling parameters above the transition point. (b) The resonance peaks of $S(\omega)$ are plotted for couplings $ \lambda^2 = 0 $ to $ \lambda^2 = 0.8 $ with steps of $ 0.1 $. The black dotted line is a guide to the eye. (c) Plot of the spectrum $S(\omega)$ for a finite temperature of $ k_BT/\omega_q = 0.1 $. As illustrated in (d), in this case the wider thermal spread of the initial resonator state (lower green solid line) compared to the ground state wave packet (lower orange dashed line) enables transitions into excited states which are localized in the wells, away from $ X =0 $. For all plots we have assumed $ \mu = 10^4 $ and $ \Gamma/\omega_q = 0.005 $.}
        \label{fig:5_spectrum_thermal}
    \end{center}
\end{figure}

As a second approach to observe the structural change in the excited state potential more directly, 
we consider in Fig.\ \ref{fig:5_spectrum_thermal}(c) the excitation spectrum at finite temperature. In this case, Eq.~\eqref{eq:SpectrumApp} must be averaged over a thermal distribution of initial states $|\phi_{0,k}\rangle$. This means that also resonator states further away from the center contribute and $S(\omega)$ probes the excited-state potential over a much larger range. For the considered temperature of $ k_BT/\omega_q = 0.1 $, the qubits are initially still in the ground state with high probability, while a large number of resonator states $\sim k_BT/\omega_r =10$ are occupied. We see that under such conditions, the triplet line splits into two distinct branches after the transition point. As illustrated in Fig.\ \ref{fig:5_spectrum_thermal}(d), these two lines correspond to the energy separations between the potential curves evaluated at $X=0$ and at the minima $X=X_{\rm min}$ of the excited state potential, respectively. Since this measurement does not require any pre-cooling of the resonator, it provides a simple way to detect first signatures of the structural change of the potential curve.

\section{Wave packet dynamics}\label{sec:Dynamics}
\begin{figure}[t]
    \begin{center}
        \includegraphics[width=\columnwidth]{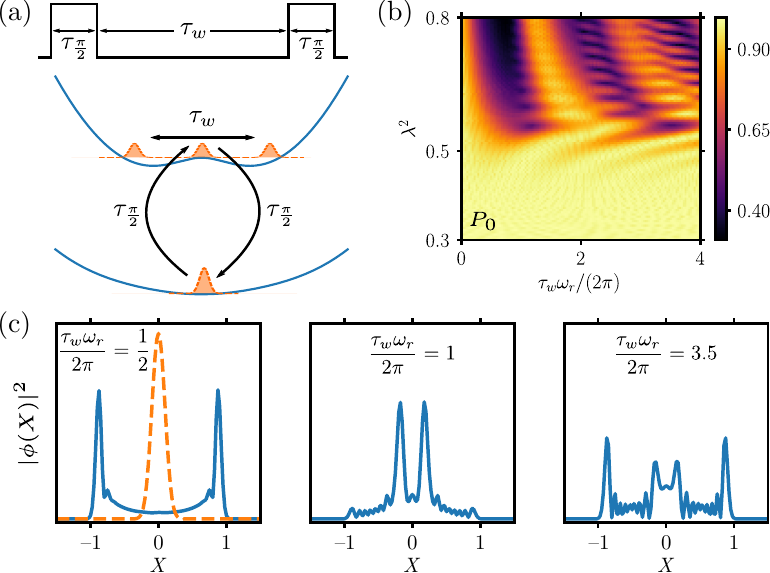}
        \caption{(a) Illustration of the Ramsey protocol for measuring the wave packet dynamics in the excited state potential. At first the system is prepared in an equal superposition between the ground- and the excited state with different adiabatic potentials. During a waiting time $ \tau_w $ the resonator wave packets evolve freely in these potentials. After a second $\pi/2$-pulse, the probability to return to the ground state, $P_0(\tau_w)$, is measured.  (b) Plot of the resulting return probability $P_0$ as a function of the free evolution time and for varying coupling strengths. (c) Snapshots of the resonator wave function in the excited state for $ \lambda^2 = 0.8 $. The initial Gaussian wave packet (orange dashed line in the first panel) is a superposition of many eigenfunctions $\phi_{1,k}$ of the double-well potential and quickly starts to spread out and swash back at a later time. Note that for all plots only the coherent dynamics given by $H_{\rm EDM}$ for $\varepsilon=0$ has been simulated. }
        \label{fig:6_ramsey_scheme}
    \end{center}
\end{figure}

In this section we discuss another technique to detect changes in the potential structure through the corresponding change in the wave packet dynamics. For this purpose we consider again a situation where the ground state potential remains stable while the first excited potential curve exhibits a symmetry-breaking transition. The basic idea is to perform a Ramsey-type measurement as depicted in Fig.~\ref{fig:6_ramsey_scheme}(a). Here the system is initially prepared in a superposition between the ground and the first excited qubit state such that the electromagnetic mode moves along two different potential curves simultaneously. Measurements of the qubit coherence can then be used to determine the overlap between  the two wave packets as a function of time. 

In Fig.~\ref{fig:6_ramsey_scheme} we illustrate this measurement protocol for the same setting as in the previous section, $N=2$ and $\varepsilon=0$. In this case, the symmetry-breaking transition occurs first in the excited potential at $\lambda=1/\sqrt{2}$. For this parameter, the ground-state potential is still approximately harmonic and active cooling methods can be applied to initialize the system in the absolute ground state $|\Psi(t=0)\rangle =|{\rm GS}\rangle\simeq |\phi_{0,0}\rangle \otimes |\chi_0(0)\rangle$. In a first step a microwave field of frequency $\omega_d$ is used to implement a $\pi/2$-rotation, which prepares the qubit state in an equal superposition between $|\chi_0(0)\rangle$ and the triplet state $|\chi_T(0)\rangle$. During this time the systems evolves according to the Hamiltonian $H(t)=H_{\rm EDM}+H_{\Omega}(t)$, where
\begin{equation}
H_{\Omega}(t) = \Omega\cos(\omega_d t + \theta)\sigma_x^1.
\end{equation}
To prepare the superposition with a fidelity of about $0.95$, we set $ \theta = 0 $, tune the frequency into resonance, $\omega_d= V_T(0)-V_{0}(0)$ and optimize the pulse time, $\tau_{\frac{\pi}{2}}\sim \pi/(2\Omega)$, for each set of parameters. Next, the system evolves freely for a  waiting time $\tau_w$ during which the wave packet propagates along two different potential curves. In a last step, a second $\pi/2$-rotation, now with an optimized phase $ \theta $, is applied, such that for $\lambda=0$ the qubits would be rotated back into the  ground state. In the interacting system, variations of the return probability $P_{0}(\tau_w) = \mathrm{Tr}\{\rho(t_f)\ket{\chi_0(0)}\bra{\chi_0(0)}\}$, where $\rho(t_f)$ is the density operator of the system at the final time $t_f=2\tau_{\frac{\pi}{2}}+\tau_w$, can be used to probe the wave packet dynamics during the free evolution time. This probability can be measured, for example, via regular dispersive readout schemes for qubits~\cite{Blais2004}.

In Fig.~\ref{fig:6_ramsey_scheme}(b) we plot the return probability $P_{0}$ as a function of $\tau_w$ and for varying coupling strengths. Note that in the numerical simulations the phase of the second $\pi/2$-pulse, $\theta$, has been adjusted for each parameter set to compensate trivial phase rotations due to a fixed energy offset between the ground and the excited state. The remaining variations then depend only on the wave packet dynamics and clearly distinguish the regime below and above the transition point. In the former case both potential curves are approximately quadratic and both wave packets remain localized around the origin. Above the transition the part of the wave packet promoted to the excited potential curve is expelled from the central region and undergoes oscillations. The resulting periodic decrease and revival of the wave packet overlap is clearly seen in the Ramsey fringes.

We verify that the observed modulation frequency in the Ramsey signal, which increases with larger $\lambda$, is consistent with the dynamics of a wave packet that is initialized at the center of the corresponding double-well potential. This confirms that the described Ramsey-protocol probes the wave packet dynamics in the excited state potential, more precisely, its overlap with the Gaussian ground state at $X=0$. To further illustrate this point, Fig.~\ref{fig:6_ramsey_scheme}(c) shows four snapshots of the actual resonator wave function during the protocol, after projecting the system on the excited triplet state $|\chi_T(0)\rangle$. Initially, at $ \tau_w = 0$, the wave packet is a Gaussian centered around $ X = 0 $. After a waiting time $\tau_w\sim \omega_r^{-1}$ most of the wave packet has propagated away from the central region, reducing the overlap with the ground state wave function. At even longer times, the reflected wave packets return, but due to the nonlinearity of the potential there is no perfect revival. This explains also the overall decay of the Ramsey fringes over a few oscillation periods. Note that in order to resolve these revivals, the qubit decoherence time $T_2$ must be longer than $\tau_w\sim \omega_r^{-1}$. Even for a resonator frequency as low as $\omega_r/(2\pi)\approx 10\,\mathrm{MHz}$, this condition can be fulfilled with realistic coherence times of $T_2=1-100\,\mu$s~\cite{Yan2016}.

\section{Implementation}\label{sec:Implementation}

In our model introduced in Sec.~\ref{sec:EDM} we have considered the coupling of multiple flux qubits to a single resonator mode. For resonant systems, $\omega_r\approx \omega_q$, and moderate couplings such a situation can be realized by using a lumped-element $LC$ resonator. In this case  the frequency of the fundamental mode, $\omega_r$, can be well separated from the next higher excitation mode with frequency $\omega_{\rm ex}$, such that even under USC conditions $\omega_{\rm ex}\gg \omega_r,\omega_q,g$~\cite{yoshihara17}. However, to observe the physics described in this work, we are interested in rather extreme ratios $\omega_q/\omega_r \gtrsim 100$, where the validity of the single-mode approximation must be evaluated in more detail.

\begin{figure}[t]
    \begin{center}
        \includegraphics[width=\columnwidth]{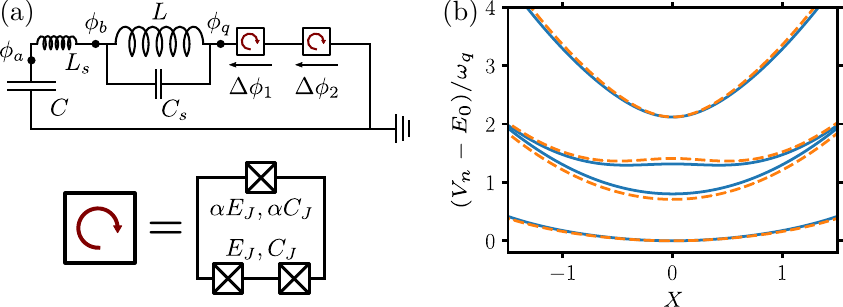}
        \caption{(a) Sketch of an extended circuit model, which includes the self-capacitance $C_s$ and the self-inductance $L_s$ of the $LC $-resonator. This can be used to estimate the influence of a higher excited resonator mode with frequency $\omega_{\rm ex}\approx 1/\sqrt{L_sC_s}$. The qubits are modelled as a superconducting loop intersected by three Josephson junctions, where the size of the upper junction is scaled by a parameter $\alpha$. (b) Comparison of the BO potentials resulting from the full two-mode circuit shown in (a) (blue solid lines) and the ideal BO potentials from the single-mode EDM (orange dashed lines). The parameters used for this plot are summarized  in Table~\ref{tab:exp_params}.Note that for the simplified single-mode model the  capacitance $C_J$ has been adjusted to have the same qubit frequency $\omega_q$ in both simulations ($ C_J \simeq 3.5\,\mathrm{fF} $ for single mode circuit).}
        \label{fig:7_cqed_imp}
    \end{center}
\end{figure}

\subsection{Two-mode circuit}
Figure~\ref{fig:7_cqed_imp}(a) shows a more realistic model for a two-qubit circuit QED system, where the self-capacitance $ C_s $ of the inductor and the self-inductance $ L_s $ of the capacitor are taken into account \cite{caloz2004}. In this case the resonator supports a second  high-frequency mode with $\omega_{\rm ex}\sim1/\sqrt{L_sC_s}$. In addition, we model each qubit by a superconducting loop with three Josephson junctions, as implemented in many experiments \cite{orlando1999,macha2014,schwarz2013,deppe2008}.

By choosing the generalized flux variables $\phi_a$ and $\phi_b$ indicated in Fig.~\ref{fig:7_cqed_imp}(a) and the phase jumps $\Delta \phi_{1,2}$ across each qubit as dynamical degrees of freedom,  the Hamiltonian for this circuit can be written as (see App.~\ref{app:2mode_circuit})
\begin{align}\label{eq:2mode_ham_full}
    H=H_{LC}+H_{\rm int}+H_q.
\end{align}
Here 
\begin{equation}\label{eq:2mode_ham_em}
\begin{split}
    H_{LC} = &\dfrac{Q_a^2}{2C} + \dfrac{Q_b^2}{2C_b} + \dfrac{(\phi_a - \phi_b)^2}{2L_s}+ \frac{\phi_b^2}{2L}
    \end{split}
\end{equation}
is the Hamiltonian for the $LC$ resonator and 
\begin{equation}\label{eq:2mode_ham_int}
\begin{split}
    H_{\rm int} = \dfrac{Q_bQ_q}{C_q} + \dfrac{\phi_q\phi_b}{L}
    \end{split}
\end{equation}
is the qubit-resonator coupling. In Eq.\ \eqref{eq:2mode_ham_em} and \eqref{eq:2mode_ham_int} the operators $Q_{a,b}$ and $Q_q$ are the conjugate charges for the flux variables $\phi_{a,b}$ and the collective qubit flux $ \phi_q = \Delta \phi_1 + \Delta \phi_2 $, respectively, and $ C_q = C_J(\alpha + 1/2) $ and $ C_b = C_sC_q/(C_q + 2C_s) $ are the relevant capacitances. Finally, $H_q$ is the Hamiltonian for the qubit degrees of freedom, which includes a correction term $\phi_q^2/(2L)$ from the inductive coupling. Its precise from is given in App.~\ref{app:2mode_circuit}, but it is not essential for following discussion.

The quadratic Hamiltonian for the electromagnetic modes can be diagonalized and written in terms of the normal mode operators $c_\pm$ as $ H_{LC} = \sum_{\eta = \pm} \omega_\eta c^{\dagger}_{\eta}c_{\eta} $. In the limit of interest, $L_s\ll L$ and $C_s\ll C$, we find that $\omega_-\approx 1/\sqrt{LC}=\omega_r$ and $\omega_+\approx1/\sqrt{L_sC_b}\gg \omega_r $. In terms of these eigenmodes, the coupling to the qubits is given by
\begin{align}\label{eq:normal_mode_coupling}
    H_{\rm int} = \sum_{\eta = \pm} \im \dfrac{ g_{Q,\eta}}{2}(c_{\eta}^{\dagger} - c_{\eta})n_q + \dfrac{ g_{\phi, \eta}}{2}(c_{\eta}^{\dagger} + c_{\eta})\varphi_q.
\end{align}
Here we have defined the dimensionless qubit variables $ \varphi_q = \phi_q/\Phi_0 $ and $ n_q = Q_q/(2e)$, where $\Phi_0$ is the reduced flux quantum.  In general, the inductive and capacitive couplings $ g_{\phi,\pm}$ and  $g_{Q,\pm}$ have a complicated dependence on all the system parameters (see App.~\ref{app:2mode_circuit}), but in the limit $\omega_-\ll \omega_+$ we obtain the approximate scaling for the low-frequency mode
\begin{equation}
\frac{g_{\phi,-}}{\omega_-} = \sqrt{\dfrac{R_Q}{\pi Z}},\qquad  \frac{g_{Q,-}}{\omega_-} = 2\sqrt{\dfrac{\pi Z}{R_Q}}\dfrac{C_b}{C_q}
\end{equation}
and
\begin{equation}
\frac{g_{\phi,+}}{\omega_+} = \sqrt{\dfrac{R_Q}{\pi Z_b}}\dfrac{L_s}{L},\qquad  \frac{g_{Q,+}}{\omega_+} = 2\sqrt{\dfrac{\pi Z_b}{R_Q}}\dfrac{C_b}{C_q}
\end{equation}
for the high-frequency mode. Here $ R_Q = h/(2e)^2 \approx 6450\,\Omega $ is the resistance quantum. We see that for an $LC$ resonator with an impedance $Z = \sqrt{L/C} \approx 50\,\Omega$, a strong, primarily inductive coupling with $g_{\phi,-}/\omega_->1$ can be achieved. Assuming $Z_b=\sqrt{L_s/C_b}\sim Z$, the coupling to the high frequency mode is still substantial in absolute numbers, but the relevant ratio $g_{\phi,+}/\omega_+$ is suppressed by a factor $L_s/L$. 

After introducing rescaled quadratures for the low-frequency mode, as in Eq.\ \eqref{eq:scaled_quadratures}, we arrive at the rescaled Hamiltonian $\tilde H=H/\omega_q$, where  
\begin{align}\label{eq:2mode_ham_P_X}
    \tilde H = \dfrac{P^2}{2\mu} +  \dfrac{X^2}{2}  + \tilde{H}_q(X, P),
\end{align}
$\omega_q$ is the characteristic qubit frequency and $ \mu = \omega_q^2/\omega_-^2 $, as above.  The Hamiltonian for the high-frequency dynamics, 
\begin{align}\label{eq:2mode_qubit_ham}
    \tilde{H}_q(X, P) = \tilde \omega_+ c^{\dagger}_+c_+ + \tilde H_{\rm int}(X, P) + \tilde H_q,
\end{align}
now also includes the high-frequency mode and depends in general also on the momentum coordinate $ P $ due to the capacitive coupling. Note that here we do not make a two-level approximation such that ${H}_q(X, P)$ still contains the exact dynamics of all high-frequency degrees of freedom, which depend parametrically on the slowly varying coordinates $X$ and $P$.

\subsection{Discussion}

In Table \ref{tab:exp_params} we present a set of parameters for a realistic setup of two qubits, which result in $ \omega_q/(2\pi) \approx 8\,\mathrm{GHz} $, $ \omega_-/(2\pi) = 50\,\mathrm{MHz} $ and  an effective mass of $\mu\approx 2.5 \times 10^4$. The coupling parameters are $g_{\phi,-}/\omega_-\approx 7.15$ and $g_{Q,-}/\omega_-\approx 0.06$. In a pure single-mode model these values translate into a coupling parameter of $ \lambda^2 \approx 0.7 $, which is well above the symmetry-breaking transition in the first excited state. Note that by varying the qubit parameters by external magnetic fluxes, this coupling could also be gradually tuned across the transition \cite{forndiaz17,Armata2017}. For the chosen values for $L_s$ and $C_s$ we obtain  $\omega_{+}/(2\pi)\approx 160$ GHz, well above the qubit energy, and $ g_{Q,+}/\omega_+ \approx 0.37$ and $ g_{\phi,+}/\omega_+ \approx 0.01 $. The value of $L_s$ is extrapolated from the parameters reported in Ref.~\cite{mckenziesell19} and could even be much lower. The assumed value of $C_s$ is within a factor of $\sim$2-3 of simulated and measured values for high-impedance coil inductors~\cite{fink2016,barzanjeh17,barzanjeh19}. For these parameters we compare in Fig.\ \ref{fig:7_cqed_imp} (b) the BO potentials obtained from the full two-mode Hamiltonian, Eq.\ \eqref{eq:2mode_qubit_ham}, to the potential of the simplified one-mode EDM used in the previous sections. The potentials are qualitatively similar with only small shifts near $ X = 0 $. Note that in our simplified model, the main limitation for $C_s$ arises from the renormalization of the qubit parameters, which partially is an artefact of modelling the distributed self- and stray capacitances of the real device in terms of a single capacitor. Leaving this aspect aside, much higher values of $C_s$ and much lower values of $\omega_+$ can be tolerated without substantially affecting the properties of the low-frequency mode.

As mentioned above, in the full model the adiabatic qubit energies obtained from of Eq.\ \eqref{eq:2mode_qubit_ham} also depends on the momentum quadrature $ P $. However, since this effect arises from the much smaller charge coupling and is proportional to $S_y$, it only induces a small second order term $\sim (g_{Q,-}/\omega_q)^2 P^2$. This correction only results in a tiny renormalization of the effective mass, which is negligible for the current set of parameters. This analysis shows that the operation of circuit QED systems in the extreme regime of ultrastrong coupling and very low frequency is experimentally feasible.  Although the current model is still an oversimplification of a real device, we expect that as long as $\omega_{\rm ex}>\omega_q$, the presence of higher excitation modes renormalizes the qubit parameters, but will not substantially change the qualitative features of the adiabatic potential curves and the measurement schemes discussed in this work.

\begin{table}[t]
        \label{tab:exp_params}
    \begin{tabular}{l|l}
        Qubit & Resonator\\
        \hline
        $ C_J = 2.21\,\mathrm{fF} $ & $ C = 79.58\,\mathrm{pF} $\\
        $ E_J/h = 336.8\,\mathrm{GHz} $ & $ L = 127.3\,\mathrm{nH} $\\
        $ \alpha = 0.74 $ & $ C_s = 1.06\,\mathrm{fF} $\\
        { } & $ L_s = 1.27\,\mathrm{nH} $
    \end{tabular}
    \caption{The set of circuit parameters for the qubits and the $LC$ resonator, which are used for the numerical simulations shown in Fig.~\ref{fig:7_cqed_imp}(b). }
\end{table}

\section{Conclusions}\label{sec:Conclusions}
In summary, we have analyzed the ultrastrong coupling of multiple superconducting flux qubits to a radio-frequency electromagnetic mode. In this regime, the dynamics of the resonator mode can be modelled as an effective particle, which moves along the adiabatic potential curves generated through the coupling to the qubits. We have shown that already for a simple two-qubit setting, the first excited potential exhibits a transition from a single- to a double-well configuration. In the limit $\omega_r\rightarrow 0$ such a transition can be used as a minimal instance to study properties of quantum phase transitions, as discussed in several previous works.  Importantly, characteristic signatures of this transition can be probed with regular spectroscopic measurements performed only on the qubits.
 
These basic findings of this work clearly demonstrate that circuit QED systems in the USC regime are not limited to tests of conventional cavity QED physics. By designing more complex potentials  and measurement schemes, this adiabatic approach can be used to access a large variety of particle-like physics and wave packet phenomena with a quasi-continuum of available states. This is very different from conventional superconducting quantum circuits, where one usually only has access to a few discrete modes. While in this work we have explicitly evaluated these capabilities for a low- frequency resonator mode, an alternative strategy to access this regime could be to work with $LC$ resonators of only slightly below $ 1\,\mathrm{GHz} $ and push the qubit frequencies to several tens of GHz. Also in this case, values of $\mu\sim 10^3-10^4$ and $\lambda\gg1$ are feasible, but different types of measurement strategies will be required.

\acknowledgments
We thank Georg Arnold for valuable feedback on circuit parameters. This work was supported by the Austrian Science Fund (FWF) through Grant No. P 31701-N27 and DK CoQuS, Grant No. W 1210, and by an ESQ Discovery Grant of the Austrian Academy of Sciences (ÖAW). J. J. Garc\' ia-Ripoll acknowledges support from AEI Project PGC2018-094792-B-I00, CSIC Research Platform PTI-001, and CAM/FEDER Project No. S2018/TCS-4342 (QUITEMAD-CM).

\appendix

\section{Adiabatic potentials}\label{app:per_thy}
In this appendix we provide approximate analytic results for the adiabatic potentials for certain limiting cases of interest.
\subsection{Double well transition of two qubit EDM}
For two qubits we can obtain the instability point of the excited state analytically through fourth-order perturbation theory. Our starting point is the qubit Hamiltonian $\tilde H_q$ given in Eq.~\eqref{eq:h_adb}. Including the bare potential $X^2/2$ for convenience it can be written as $\tilde{H}_{q} = \tilde{H}_0 + \tilde{H}_1$,
where
\begin{align}
    \tilde{H}_0 &= \dfrac{X^2}{2} + S_z + (1 + \varepsilon)\lambda^2 S_x^2,\\
    \tilde{H}_1 &= \sqrt{2}\lambda XS_x.
\end{align}
In the case of two qubits we can diagonalize $ \tilde{H}_0 $ analytically because the triplet state $ | m_z = 0 \rangle \equiv | T \rangle $, is decoupled from the other two states $ | m_z = \pm 1 \rangle \equiv | \pm 1 \rangle $. The eigenenergies and eigenstates of $ \tilde{H}_0 $ are
\begin{align}
    \tilde{E}_T^{(0)} &=  \dfrac{X^2}{2} + \bar{\varepsilon}\lambda^2, \qquad | \chi_T \rangle = | T \rangle,
\end{align}
and
\begin{align}
    \tilde{E}_{\pm}^{(0)} &=   \dfrac{1}{2}\left( X^2 + \bar{\varepsilon}\lambda^2 \pm \sqrt{4 + \bar{\varepsilon}^2\lambda^4} \right),\\
    | \chi_{\pm} \rangle &= N_{\pm}^{-1} \left[ \left( 2\bar{\varepsilon}^{-1}\lambda^{-2} \pm \sqrt{1 + 4\bar{\varepsilon}^{-2}\lambda^{-4}} \right) | 1 \rangle + | -1 \rangle \right].
\end{align}
Here we have introduced the abbreviation $ \bar{\varepsilon} = 1 + \varepsilon $ and the normalization factor $ N_{\pm} = \sqrt{1 + \left( \sqrt{1 + 4\bar{\varepsilon}^{-2}\lambda^{-4}} \pm 2\bar{\varepsilon}^{-1}\lambda^{-2} \right)^2} $. The singlet state $ \ket{S} $ is decoupled from the resonator and thus has the energy $ \tilde{E}_S = X^2/2 $, independent of $ \lambda $. This leads to the splitting of the singlet and triplet states given in Eq.~\eqref{eq:DeltaE}.

To obtain the instability point for the excited state $ \ket{\chi_T} $ we calculate the perturbative corrections resulting from $ \tilde{H}_1 $ up to fourth order. The odd order perturbative corrections vanish since the semiclassical Hamiltonian is symmetric under the transformation $ X \to -X,\, S_x \to -S_x $. Thus, the energy of $ | \psi_T \rangle $ to fourth order is given by
\begin{align}\label{eq:E4}
    \tilde{E}_T = \tilde{E}_T^{(0)} + \tilde{E}_T^{(2)} - \tilde{E}_T^{(2)} \sum_{\eta = \pm} \dfrac{| \langle \chi_{\eta} | \tilde{H}_1 | \chi_{T} \rangle |^2}{( \tilde{E}_T^{(0)} - \tilde{E}_{\eta}^{(0)} )^2},
\end{align}
where 
\begin{align}
    \tilde{E}_T^{(2)} = \sum_{\eta = \pm} \dfrac{| \langle \chi_{\eta} | \tilde{H}_1 | \chi_{T} \rangle |^2}{\tilde{E}_T^{(0)} - \tilde{E}_{\eta}^{(0)}}
\end{align}
is the second order correction. Note that other than the last term in Eq.~\eqref{eq:E4}, all other fourth-order correction terms vanish. From the zeroth-order results form above we find
\begin{align}
    \tilde{E}_T^{(0)} - \tilde{E}_{\pm}^{(0)}
    &= \dfrac{1}{2} \left( \bar{\varepsilon}\lambda^2 \mp \sqrt{4 + \bar{\varepsilon}^2\lambda^4} \right),\\
    \langle \chi_{\pm} | \tilde{H}_1 | \chi_{T} \rangle
    &= \dfrac{\lambda X}{N_{\pm}} \left( 1 + 2\bar{\varepsilon}^{-1}\lambda^{-2} \pm \sqrt{1 + 4\bar{\varepsilon}^{-2}\lambda^{-4}} \right).
\end{align}
Therefore, in total we obtain
\begin{align}
    \tilde{E}_T^{(2)} &= -2X^2\bar{\varepsilon}\lambda^4,\\
    \tilde{E}_T^{(4)}& = 4X^4\bar{\varepsilon}\lambda^6(1 + \bar{\varepsilon}^2\lambda^4),
\end{align}
such that
\begin{align}
    \tilde{E}_T = \lambda^2 + \dfrac{X^2}{2}(1 - 4\bar{\varepsilon}\lambda^4) + 4X^4\bar{\varepsilon}\lambda^6 (1 + \bar{\varepsilon}^2\lambda^4).
\end{align}
We see that $ \tilde{E}_T $ transforms from a single well to a double well when the quadratic term in $ X $ changes sign, i.e.,\ at
\begin{align}
    4\bar{\varepsilon}\lambda^4 = 1 \Leftrightarrow \dfrac{g^2}{\omega_q\omega_r} = \dfrac{1}{2\sqrt{1 + \varepsilon}}.
\end{align}
The positions of the new minima above $ \lambda^2 > 1/(2\sqrt{1 + \varepsilon}) $ are not predicted accurately by the above perturbative model since at the minimum the condition $ \lambda X \ll 1 $ is not respected anymore.

\subsection{Structure of the ground- and first excited-state potentials for $ \lambda \gg 1 $}
In Sec.~\ref{subsec:General} we showed that in the limit $ \lambda \gg 1 $ the effective potentials are to a first approximation given  by
\begin{align}
    \tilde  V_{s,m_x}^{(0)}(X) = \dfrac{1}{2}(X + \sqrt{2}\lambda m_x)^2 + \varepsilon\lambda^2m_x^2.
\end{align}
We calculate the effect of the free qubit Hamiltonian $ S_z $ on these potentials at the minima $ X = -\sqrt{2}\lambda m_x $ and at the points where two neighbouring minima meet $ X = -(1 + \varepsilon)(2m_x + 1)\lambda/\sqrt{2} $. The second order perturbative correction to the potentials is given by
\begin{align}
    \tilde{V}_{s,m_x}^{(2)}(X) = \sum_{m_x' \neq m_x} \dfrac{|\bra{m_x'} S_z \ket{m_x}|^2}{\tilde V_{s,m_x}(X) - \tilde V_{s,m_x'}(X)},
\end{align}
which holds for $ X $ away from the degeneracy points. At the minima $ X = -\sqrt{2}\lambda m_x $ we obtain
\begin{align}
    &\tilde{V}_{s,m_x}^{(2)}(-\sqrt{2}\lambda m_x) = \nonumber\\
   & \dfrac{-1}{4\lambda^2}\left( \dfrac{s_+^2(m_x)}{1 + (2m_x + 1)\varepsilon} + \dfrac{s_-^2(m_x)}{1 - (2m_x - 1)\varepsilon} \right)\\
    &\approx \dfrac{(1 - 3\varepsilon)m_x^2 - (1 - \varepsilon)s(s + 1)}{2\lambda^2}\nonumber,
\end{align}
where $ s_{\pm}(m_x) = \sqrt{s(s + 1) - m_x(m_x \pm 1)} $, and to obtain the simplified expression we have assumed weak interactions $ |\varepsilon(2m_x \pm 1)| \ll 1 $. Thus, for weakly interacting systems the $ m_x = 0 $ state emerges as the ground state.
Note that this semiclassical result agrees with the predictions from a strong-coupling perturbation theory of the full model for $\varepsilon=0$~\cite{jaako2016}. 

At the degeneracy points we can diagonalize the two by two matrix describing the two spin states $ m_x$ and  $m_x + 1 $. The reduced Hamiltonian is given by
\begin{align}
    \tilde H_{X = -\lambda(1 + \varepsilon)(2m_x + 1)/\sqrt{2}} = \dfrac{\xi}{4} \mathrm{I}_{2\times 2} + \dfrac{s_+(m_x)}{2}\sigma_x,
\end{align}
where $ \xi = \lambda^2(1 + \varepsilon)\left[ 1 + (2m_x + 1)^2\varepsilon \right] $ and $ \mathrm{I}_{2\times 2} $ is the $ 2\times 2 $ identity matrix. The eigenvalues are simply
\begin{align}
    \tilde{E}_{\pm} = \dfrac{1}{2}\left( \dfrac{\xi}{2} \pm s_+(m_x) \right).
\end{align}
Thus, the splitting between the originally degenerate states is $ \tilde{\Delta}_{m_x,m_x + 1} = s_+(m_x) $.

\section{Two-mode circuit QED Hamiltonian}\label{app:2mode_circuit}
In this appendix we present additional details about the derivation of the full Hamiltonian for the circuit depicted in Fig.~\ref{fig:7_cqed_imp}(a). We follow the standard quantization approach~\cite{vool2016} and define the generalized flux variables
\begin{equation} 
\phi_{a,b} (t)=\int_{-\infty}^t \id s \, V_{a,b}(s),
\end{equation}
where $V_{a,b}(t)$ is the voltage at the nodes indicated in Fig.~\ref{fig:7_cqed_imp}(a). The qubits are described by the phase variables $\varphi_{1,2}=\Delta \phi_{1,2}/\Phi_0$, which represent the phase difference across the upper junction with Josephson energy $\alpha E_J$. Depending on the flux qubit design, there can be additional internal dynamical degrees of freedom. For the considered three-junction design, this  additional dynamical variable is the generalized flux $\phi_{-,k}$ between the two junctions in the lower arm. 

The classical dynamics of this circuit is then described by the Lagrangian
\begin{align}
    \mathcal{L} &= \frac{C\dot \phi_a^2}{2} + \frac{C_s(\dot \phi_b-\dot\phi_q)^2}{2}- \dfrac{(\phi_a - \phi_b)^2}{2L_s} - \dfrac{(\phi_q - \phi_b)^2}{2L}\nonumber\\
    & + \sum_k \bigg[ \frac{C_q\Delta \dot \phi_k^2}{2} + \frac{C_-\dot \phi_{-,k}^2}{2}+ \alpha E_J\cos\left( \dfrac{\Delta\phi_{k} + \Phi_e}{\Phi_0} \right)\nonumber\\
    &+ 2E_J \cos\left(\dfrac{\Delta\phi_{k}}{2\Phi_0}\right)\cos\left(\dfrac{\phi_{-,k}}{2\Phi_0}\right) \bigg],
\end{align}
where the effective capacitances are  $ C_q = C_J(\alpha + 1/2) $ and $ C_- = C_J/2 $ and $ \phi_q = \sum_k \Delta\phi_k $, and $ \Phi_e $ is an external magnetic flux threading the loop formed by the three junctions of the qubits. By introducing the canonical charges $Q_\eta=\partial \mathcal{L}/\partial \dot \phi_\eta$ for all flux variables and performing a Legendre transformation, we obtain the Hamiltonian given in Eq. \eqref{eq:2mode_ham_full} with  
\begin{align}
       &H_{q} = \sum_k \bigg[ \dfrac{Q_{k}^2}{2C_q} + \dfrac{Q_{-,k}^2}{2C_-} + \sum_{l}\dfrac{\Delta\phi_{k}\Delta\phi_{l}}{2L}\\
    &\;- 2E_J\cos\left(\dfrac{\Delta\phi_{k}}{2\Phi_0}\right)\cos\left(\dfrac{\phi_{-,k}}{2\Phi_0}\right) -\alpha E_J\cos\left( \dfrac{\Delta\phi_{k} + \Phi_e}{\Phi_0} \right) \bigg]\nonumber.
\end{align}
By diagonalizing the harmonic Hamiltonian $ H_{LC} $ given in Eq.\ \eqref{eq:2mode_ham_em} we obtain the mode frequencies
\begin{align}
     \omega_{\pm }^2 = \dfrac{1}{2}\left(\omega_a^2 + \omega_b^2 \pm \sqrt{(\omega_a^2 - \omega_b^2)^2 + 4g_{ab}^2\omega_a\omega_b}\right),
  \end{align}
with the bare frequencies $ \omega_a = 1/\sqrt{L_sC} $, $ \omega_b = 1/\sqrt{L_bC_b} $ and the coupling $ g_{ab} = \sqrt{Z_b/Z_a}\omega_a $. Using these normal modes, we obtain the coupling Hamiltonian $H_{\rm int}$ given in Eq.\ \eqref{eq:normal_mode_coupling}, where the couplings are 
\begin{align}
    &g_{Q,+} = 2\sqrt{\dfrac{\pi Z_b}{R_Q}}\dfrac{C_b}{C_q}\sqrt{\dfrac{\omega_+}{\omega_b}}\sin( \xi)\omega_b,\\
    &g_{Q,-} = 2\sqrt{\dfrac{\pi Z_b}{R_Q}}\dfrac{C_b}{C_q}\sqrt{\dfrac{\omega_-}{\omega_b}}\cos( \xi)\omega_b,\\
    &g_{\phi,+} = \sqrt{\dfrac{R_Q}{\pi Z_b}}\dfrac{L_b}{L}\sqrt{\dfrac{\omega_b}{\omega_+}}\sin( \xi)\omega_b,\\
    &g_{\phi,-} = \sqrt{\dfrac{R_Q}{\pi Z_b}}\dfrac{L_b}{L}\sqrt{\dfrac{\omega_b}{\omega_-}}\cos(\xi)\omega_b.
\end{align}
In these expressions the mixing angle $ \xi $ is given by
\begin{equation}
    \tan(2 \xi)= -\frac{2g_{ab}\sqrt{\omega_a\omega_b}}{\omega_a^2 - \omega_b^2}.
\end{equation}
For $ C_s, L_s \ll C, L $, respectively, we can identify $ \omega_- $ as a low-frequency mode with $ \omega_- \simeq \omega_r $. Using the rescaled quadratures of Eq.\ \eqref{eq:scaled_quadratures} we can write the total Hamiltonian as in Eq.\ \eqref{eq:2mode_ham_P_X}.

\end{document}